\documentclass[twocolumn,prl]{revtex4}
\usepackage{graphicx}
\usepackage{dcolumn}
\usepackage{amsmath}
\usepackage{bm}

\begin{document}
\title{
Is there something of the MCT in orientationally
disordered crystals ?}
\author{F.~Affouard}
\author{M.~Descamps}
\affiliation{
Laboratoire de Dynamique et Structure des Mat\'eriaux Mol\'eculaires,\\
CNRS ESA 8024,
Universit\'e Lille I,\\
59655 Villeneuve d'Ascq Cedex France}
\date{\today}

\begin{abstract}
Molecular Dynamics simulations have been performed on the 
orientationally disordered crystal chloroadamantane:
a model system where dynamics are almost completely 
controlled by rotations.
A critical temperature $T_{c} \simeq 225 $ K
as predicted by the Mode Coupling Theory can be clearly determined
both in the $\alpha$ and $\beta$ dynamical regimes.
This investigation also
shows the existence of a second remarkable dynamical crossover
at the temperature $T_{x} > T_{c}$ consistent with a previous
NMR and MD study~\cite{Affouard_epl}.
This allows us to confirm clearly the existence 
of a 'landscape-influenced' regime 
occurring in the temperature range 
$[T_{c}-T_{x}]$ as recently proposed~\cite{Sastry,Sastry2}.
\end{abstract}

\maketitle

In recent years, a great deal of work has been done to
obtain a fundamental understanding of the glassy state
and the glass formation.
Despite these studies, no well-accepted theory
has emerged so far~\cite{Ediger}.
At present, the only theory
which gives precise predictions
of the supercooled liquids dynamics is 
the Mode Coupling Theory (MCT)~\cite{Goetze_MCT}.
The idealized version
of this theory
particularly predicts
an ergodic to nonergodic transition
at the critical temperature $T_{c}$ which is
usually depicted as an
\emph{ideal-glass transition}. 
However, experimentally, no sharp transition is observed at $T_{c}$
and 
the real glass formation only occurs at the
calorimetric temperature $T_{g} <  T_{c}$.
It is suspected that some not so well understood processes \emph{i.e}
thermally activated hopping restore ergodicity
below $T_{c}$ ('landscape-dominated' regime).
During the past ten years, 
a variety of experiments~\cite{Winkler} 
(time-of-flight neutron, 
Raman,
depolarized light scattering, dielectric spectroscopy, optical Kerr effect) 
and Molecular Dynamics (MD) computer simulations 
have been made to validate MCT.
It has been particularly shown
that MCT gives an outstanding good description 
of simple 
liquids (binary Lennard-Jones, hard sphere) while 
for molecular liquids where both translational
and orientational degrees of freedom (TDOF, ODOF) coexist, 
the predictions of the MCT
seem only in fair agreement
with the experimental or the simulation data.
This raises particularly the question,
which remains a matter of debate
at present,
of the relative importance of 
rotations and translations in molecular liquids during the 
dynamical slowing down. 

One class of compounds offers attractive possibilities 
to focus mainly on the role of the ODOF
and provides a valuable
alternative to
the structural molecular liquid glass-formers.
Indeed,
some molecular crystals show a high temperature \emph{plastic} phase
in which the average position of the centers of mass are
ordered on a lattice while the orientations are
dynamically disordered.
Some of them, called
\emph{glassy crystals}~\cite{Suga}, 
such as cyanoadamantane~\cite{Descamps4}, ethanol~\cite{Criado,Benkhof}
 or 
cyclooctanol~\cite{Brand},
can be deeply supercooled 
and present many properties characteristic of
the conventional molecular liquid glasses such as a step
in the specific heat at the glass transition or a non-Arrhenius behaviour
of the relaxation times.

In the present Letter, we 
discuss precisely, by means of a case study of chloroadamantane,
the validity of MCT 
in orientationally disordered crystals.
Indeed, 
although many experimental facts have been collected for this class 
of compounds in 
the deep supercooled domain, 
we have to emphasize that it really exist very few
informations of dynamics far above $T_{g}$. This particularly 
raises the question 
if MCT could be valid for such systems where dynamics 
are mainly controlled by rotations.
One have to mention a numerical investigation made by Renner \emph{et al.}
in~\cite{Renner} where a system of colliding hard needles distributed
on a lattice, \emph{i.e} a simple model of plastic crystal,
has shown a rotational dynamical decoupling.
Stimulated by this study and by
the various predictions of MCT, we have chosen to study
the chloroadamantane $\mathrm{C_{10}H_{15}Cl}$ (noted Cla in the following) 
molecular crystal. This compound belongs to 
the substituted adamantane family
which presents excellent experimental candidates~\cite{Descamps4}.
Very recently,
by means of NMR experiment and MD computer simulation,
we have shown that this system
exhibits a dynamical crossover
transition at $T_{x} \simeq 350 $ K in the pico-nanosecond
regime~\cite{Affouard_epl}. 
It was interpreted as an indication of a 
a change 
of the energy landscape topography
as it has been recently proposed~\cite{Sastry}.
In addition, this crossover was suspected to be the
analogue of the Goldstein
crossing temperature between quasi-free diffusion and activated regime
predicted in liquids.
Two-step ($\alpha - \beta$) relaxations was also observed to emerge
but there were some indications that
$T_{x}$ probably did not
correspond to the $T_{c}$ of the MCT as developed for
structural glass-formers. 

MD simulation of orientationally disordered molecular crystal
Cla have been performed
at different temperatures from
$T =$ 220 to 500 K. 
Owing to the very long MD runs, equilibrium of the system
could not be obtained below $T \simeq 220 $ K.
A simple model was used which is 
completely described in~\cite{Affouard_AIP,Affouard_cyano},
so 
only the essential details are given in~\cite{Affouard_cladm_model}. 
\begin{figure}[h]
\includegraphics[width=6.25cm]{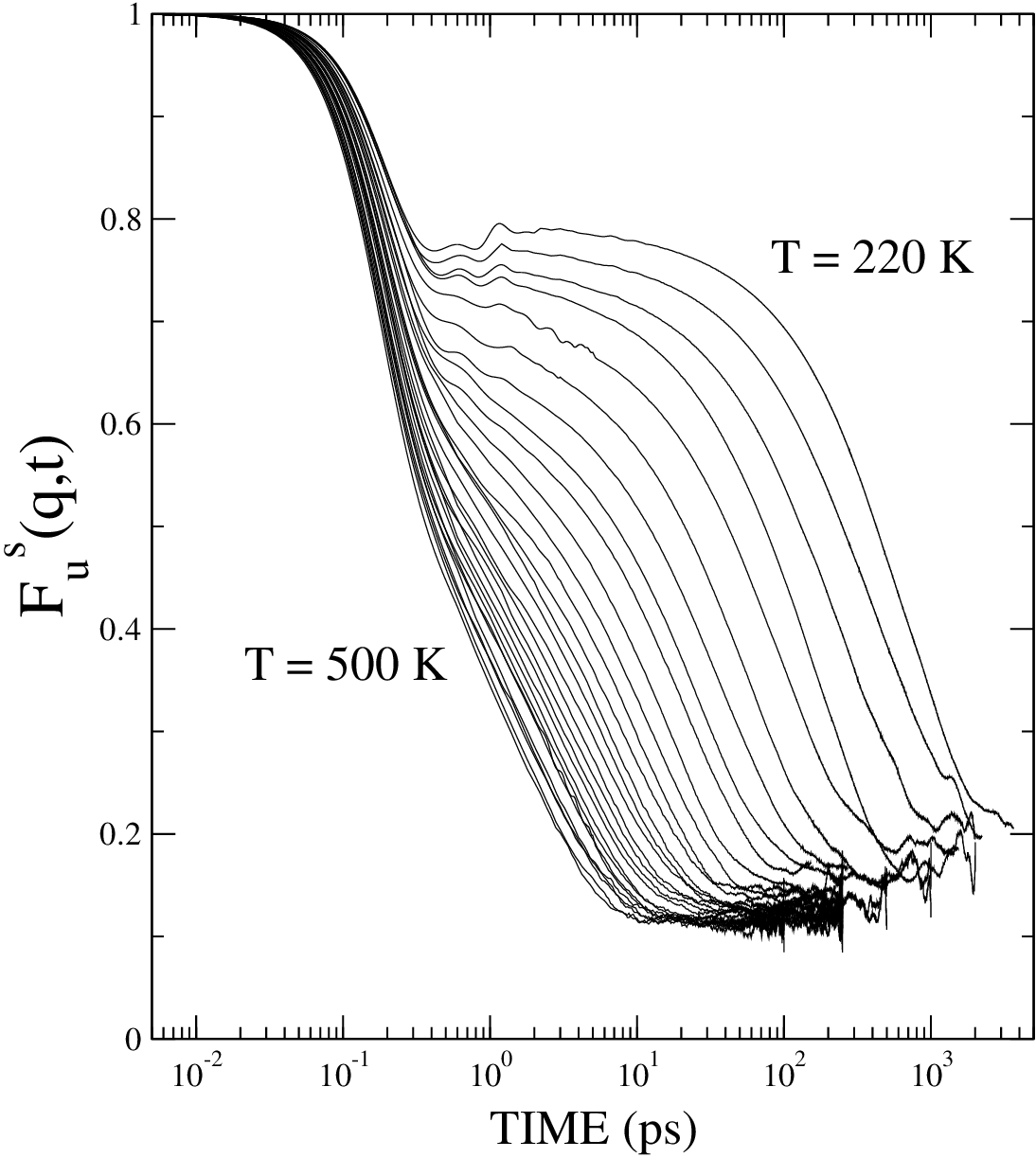}
\caption{\small
Orientational self intermediate 
scattering function $F^{s}_{u}(\vec{q},t)$
as function
of time $t$ for wave vector $q = 3.06 \ \mathrm{\AA^{-1}}$.
$F^{s}_{u}(\vec{q},t)$ is defined as 
$
\sum_{i=1}^{N}\sum_{a=1}^{N_{a}}
\left \langle
\exp[ i\vec{q}.(\vec{u}_{i,a}(t) - \vec{u}_{i,a}(0))]
\right \rangle
$
where $\vec{u}_{i,a} = \vec{R}_{i,a} - \vec{R}^{c.m}_{i}$.
$\vec{R}^{c.m}_{i}$ is the center of mass of molecule $i$ and 
$\vec{R}_{i,a}$ denotes the position of site
$a$ 
in each molecules ($N_{a}=2$ in the present study).
}
\label{figure1}
\end{figure}

\begin{figure}[h]
\includegraphics[width=6.25cm]{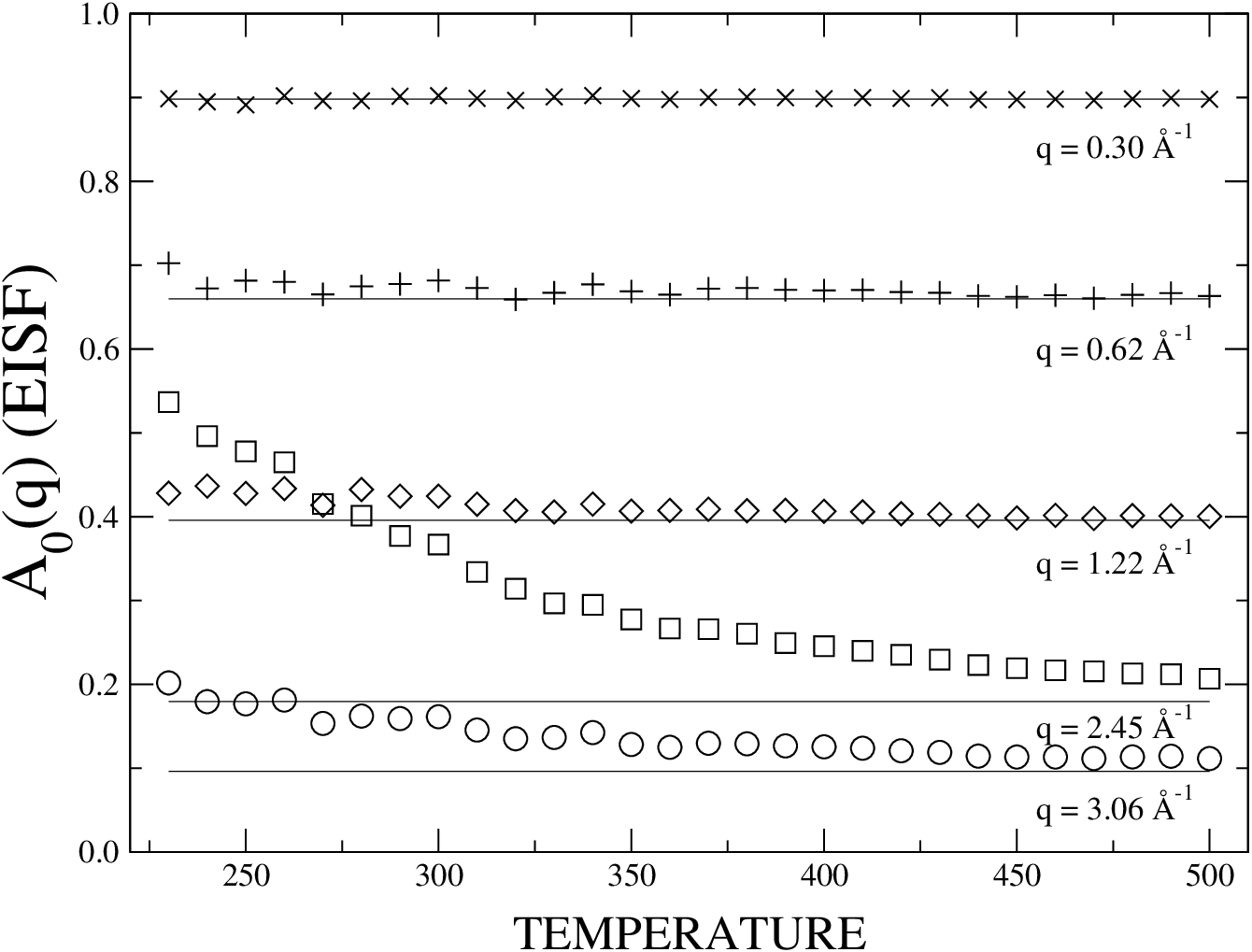}
\caption{\protect\small
Elastic incoherent structure factor (EISF) as function
of temperature for arbitrary wave vector $q = 3.06, 2.45, 1.22, 0.62 $ and
$0.30 \ \mathrm{\AA^{-1}}$.
For a rotational diffusion model (solid lines),
the EISF for one scattering atom moving on a sphere
of radius $a$ can be written:
$A_{0}(q) \sim j_{0}(q.a)$
where $j_{0}$ is the 0-order spherical Bessel function
~\cite{Sherwood}.
}
\label{figure2}
\end{figure}
Fig.~\ref{figure1} shows
the orientational self intermediate scattering function $F^{s}_{u}(q,t)$,
for arbitrary wave vector $q = 3.06 \mathrm{\AA^{-1}}$ ($\vec{q} = (q,0,0)$), 
as defined by Lewis \emph{et al.} in~\cite{Lewis2}
where only the ODOF are taken into account.
Clearly, lowering
the temperature,
as already observed for the orientational correlation functions
$C_{l}(t)=
\frac{1}{N}\sum_{i=1,N}
\left \langle
P_{l} \left(
\vec{\mu_{i}}(t).\vec{\mu_{i}}(0)
\right)
\right \rangle
$ in~\cite{Affouard_epl},
the rotator phase of Cla exhibits a two-step relaxation.
At intermediate times, $F^{s}_{u}(q,t)$
shows a plateau-like region which reveals 
the existence of an \emph{orientational cage effect} 
\emph{i.e} the rotational analogue of 
the translational cage effect observed in liquids. 
This transient regime is followed by a slow decay
to a non zero value 
called 
elastic incoherent structure factor (EISF) 
and noted $A_{0}(q) = \lim_{t \rightarrow \infty}F^{s}_{u}(q,t)$
~\cite{Sherwood}
which gives information 
on the time-average orientational geometry of the molecular motions.
$A_{0}(q)$ 
is displayed in the figure~\ref{figure2}
for different wave vectors and exhibits an unusual temperature dependence
which confirms, as already seen in~\cite{Affouard_epl},
the existence of a change of the rotational dynamics. 
Indeed, at high temperature,
an isotropic diffusion rotation model gives a good
description of dynamics. Lowering the temperature,
discrepancies of this model are clearly seen in the figure~\ref{figure2}
and a jump-like between preferred orientations
type of motion are to be assumed.

In the following, we try to describe our data in the MCT framework.
The idealized version~\cite{Goetze_MCT} of this theory predicts 
a two-step relaxation scenario (fast $\beta$, slow $\alpha$) of all the 
time dependent correlators $\phi(q,t)$. MCT particularly states 
the following points:
At short time, $\phi(q,t)$ decays to a plateau value,
the so-called nonergodicity parameter 
classically noted $f^{c}_{q}$. The dynamical regime associated 
to this plateau is called $\beta$. It is centered around 
the rescaling time $t_{\sigma}$ which is given by 
$t_{\sigma} = t_{0}|\sigma|^{-1/2a}$
where $t_{0}$ is a characteristic microscopic time and 
$\sigma$ is proportional 
to $|T -T_{c}|$. 
Above $T_{c}$,  
the late $\beta$ regime or the early $\alpha$ relaxation
is described by the following power law (going beyond first order):
\begin{equation}
\phi(q,t) = f^{c}_{q} - h^{(1)}_{q}.(t/\tau)^{b} 
+ h^{(2)}_{q}.(t/\tau)^{2b} \label{equation1}
\end{equation}
where 
the first two terms correspond to the classical von Schweidler law,
the last term is a second order correction and 
$\tau=t_{0}|\sigma|^{-\gamma}$ with $\gamma = 1/2a + 1/2b$.
At long time,
MCT also predicts that the previous fast regime 
is followed by a slow relaxation, called $\alpha$, 
with the characteristic time $\tau$. 
Furthermore, 
parameters $a$ and $b$ are 
temperature and $q$ independent and related via 
$\Gamma^{2}(1-a)/\Gamma(1-2a) = \Gamma^{2}(1+b)/\Gamma(1+2b)$
where $\Gamma(x)$ is the gamma function. 
\begin{figure}[h]
\includegraphics[width=6.25cm]{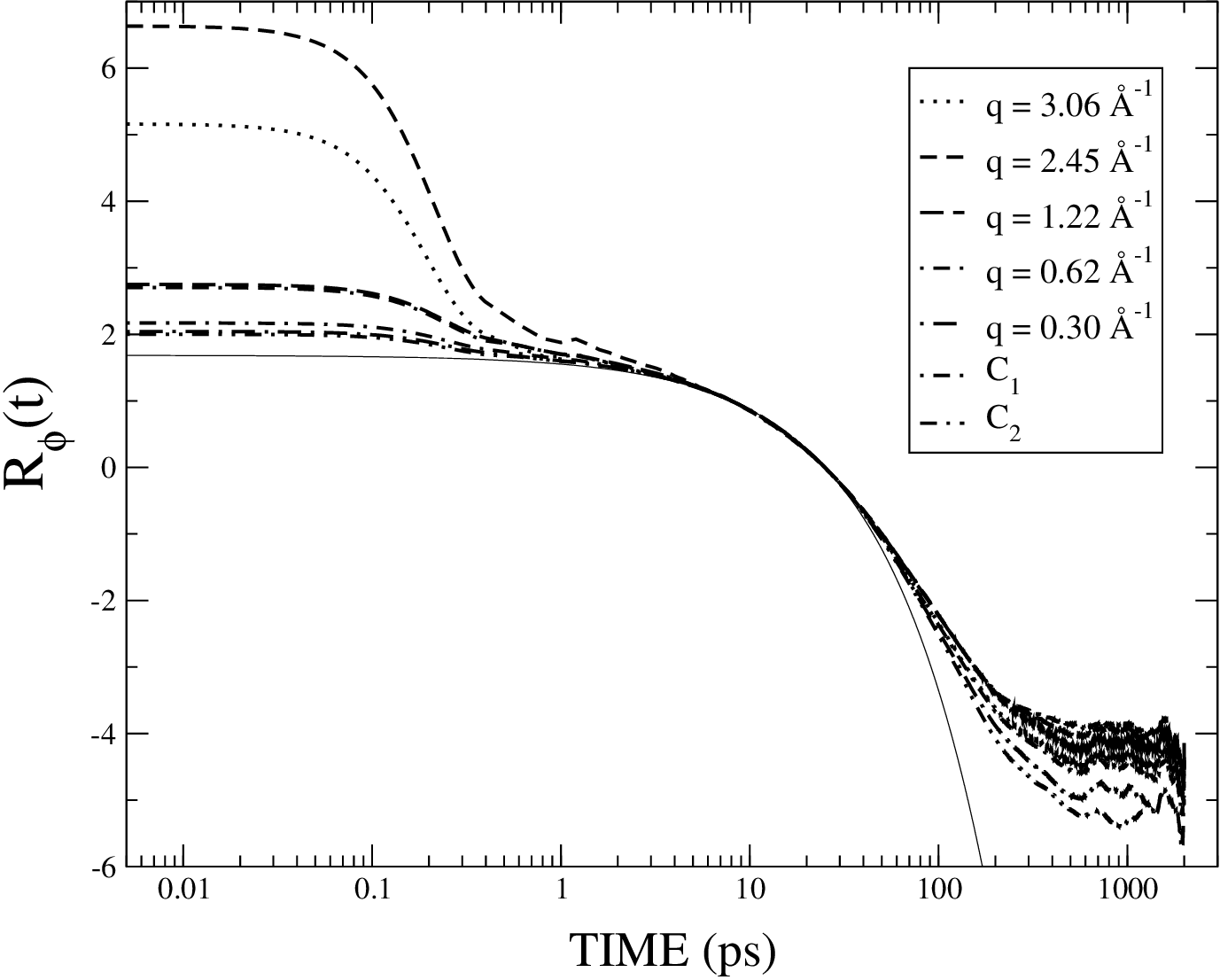}
\caption{\protect\small
Ratio $R_{\phi}(t) = (\phi(t)-\phi(t_{1}))/(\phi(t_{2})-\phi(t_{1}))$
where $(t_{1},t_{2})$ are two different arbitrary times in the
$\beta$ regime
at $T = 260 $ K.
The following correlation functions have been used: $\phi(t)
 = F^{s}_{u}(q,t)$ for wave vectors $q = 3.06, 2.45, 1.22, 0.62 $ and
$ 0.30 \ \mathrm{\AA^{-1}}$ and $\phi(t) = C_{l}(t)$ for $l=1,2$
angular correlation functions (see~\cite{Affouard_epl}).
Master curve (solid line) obtained by the von Schweidler law with
the exponent $b=0.79$ is also displayed.
}
\label{figure3}
\end{figure}
In order to verify the validity of those scaling laws, we first tried
to fix the exponent $b$ 
using the factorization theorem in the $\beta$ regime 
where hopping processes are not supposed to be dominant. 
The ratio
$R_{\phi}(t) = (\phi(t)-\phi(t_{1}))/(\phi(t_{2})-\phi(t_{1}))$,
where $(t_{1},t_{2})$ are two different arbitrary times in the
$\beta$ regime, has been calculated
at different temperatures
and for different 
correlators $\phi$ ($F^{s}_{u}(q,t)$ for several
wave vectors and $C_{l}(t)$ for $l=1,2$ as defined
in~\cite{Affouard_epl}). 
Results are displayed in the figure~\ref{figure3}. Clearly,
it exists a time domain where
the different correlation functions
collapse onto a master curve in the $\beta$ regime
as it is predicted by MCT. 
Assuming the correlator function $\phi$ departs from the plateau
with a von Schweidler law, it is possible to show 
that the master curve depends only on the exponent $b$.
Using a fitting procedure 
performed at several different temperatures
for all correlators mentioned previously,
we obtained the best results 
for $b=0.79$ which corresponds to 
$a=0.36$ and $\gamma=2.02$. Then, fixing these values,
we performed individual fit of the different correlators
in the late $\beta$ regime using the equation~\ref{equation1}
with free parameters $f^{c}_{q}$, $\tilde{h}^{(1)} = h^{(1)}(q).\tau^{-b}$ and 
$\tilde{h}^{(2)} = h^{(2)}(q).\tau^{-2b}$.
Only the temperatures below 330 K where the plateau region 
is clearly defined were considered.
The total prefactor $\tilde{h}^{(1)}$ is displayed in the figure~\ref{figure4}.
$\tilde{h}^{(2)}$ is not shown in the present Letter but it exhibits the
same temperature dependence.
A temperature dependence is clearly found which extrapolates
to zero at the critical temperature $T_{c} \simeq 225 \pm 8 $ K
for the different correlators. A deviation to the linear evolution
is observed close to $T_{c}$ which could be attributed
to the occurence of hopping processes. The nonergodicity parameter $f^{c}_{q}$
show a relatively smooth evolution as function of temperature but
we have no clear evidence of a cusp at $T_{c}$
due to the lack of data at low temperature.

\begin{figure}[h]
\includegraphics[width=6.25cm]{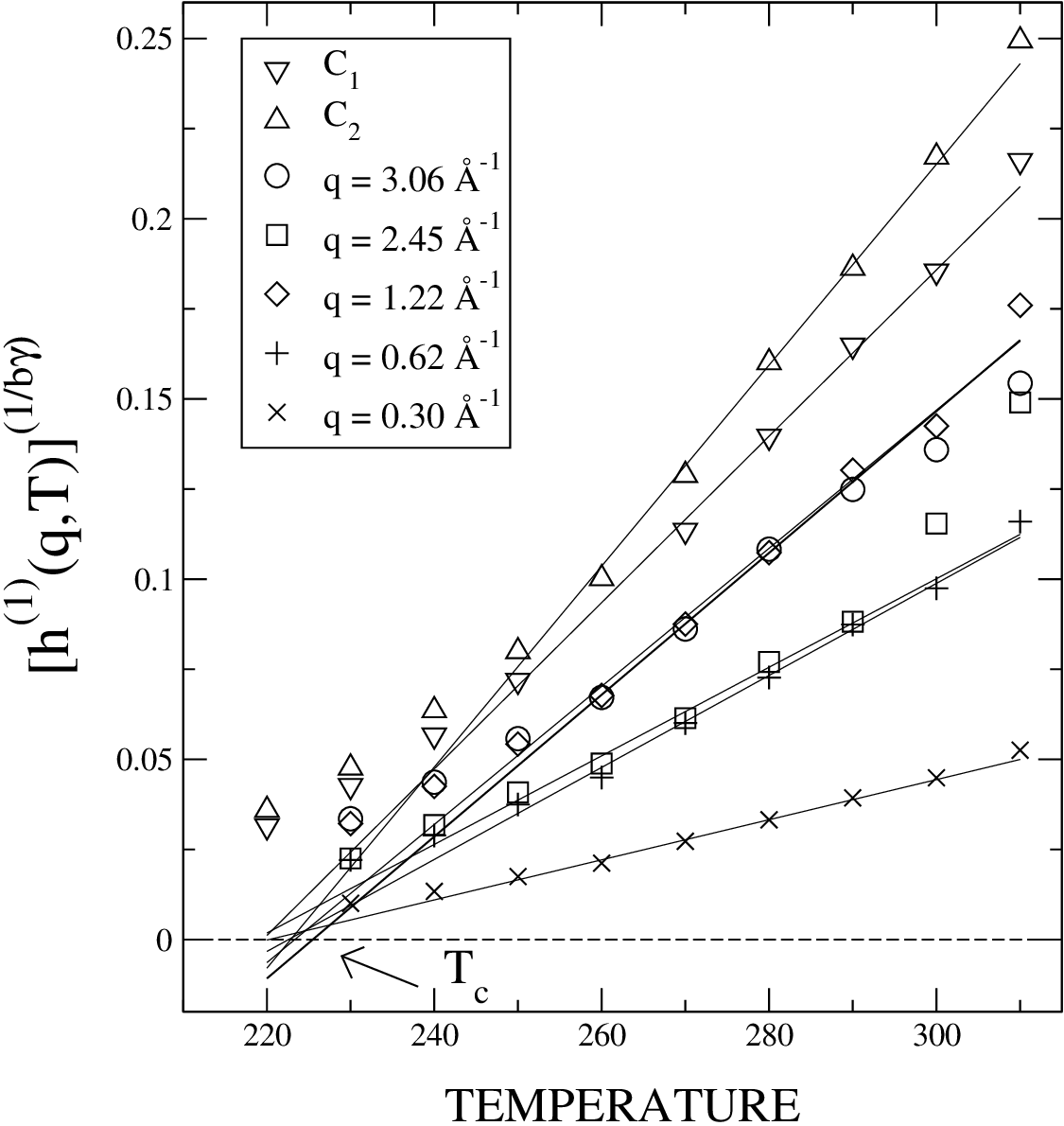}
\caption{\protect\small
Total prefactor $[\tilde{h}^{(1)}]^{1/b.\gamma}$
as function
of temperature for all investigated correlators (see Fig.~\ref{figure3}).
Fitting procedure reveals a
linear dependence (solid lines) of the total prefactor which extrapolates
to zero at the critical temperature $T_{c} \simeq 225 \pm 8 $ K for all
correlators.
}
\label{figure4}
\end{figure}
\begin{figure}[h]
\includegraphics[width=6.25cm]{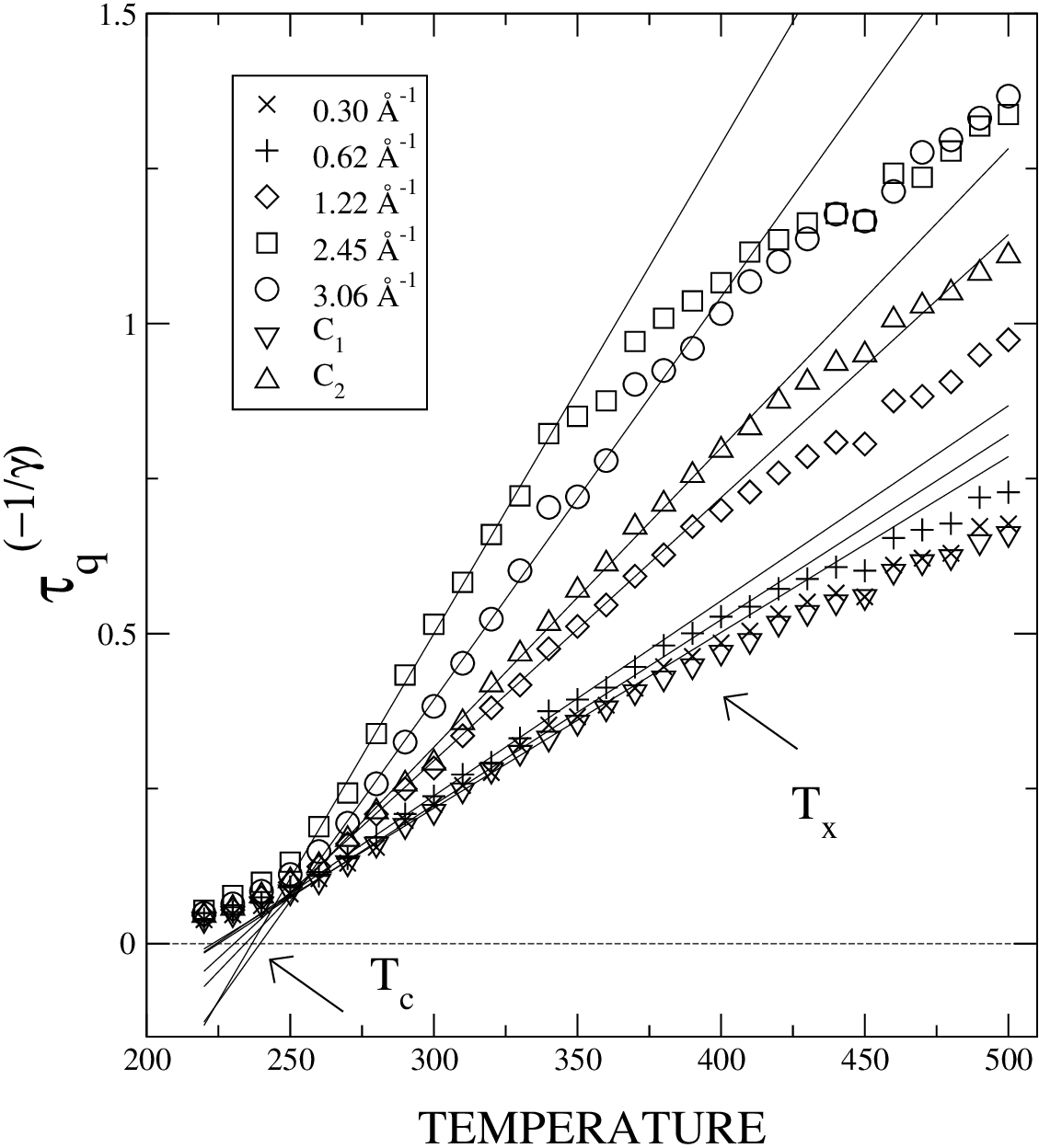}
\caption{\protect\small
$\alpha$ relaxation time $\tau^{-1/\gamma}_{q}$
as function
of temperature for all investigated correlators (see Fig.~\ref{figure3}).
The exponent $\gamma = 2.02$ has been extracted from the MCT analysis 
in the $\beta$ regime.
Solid lines indicate MCT-fit using the linear law 
$\tau^{-1/\gamma}_{q} \sim (T-T_{c})$. Deviation from 
the MCT prediction is found at $T > T_{x} \simeq 350 $ K
where $\tau_{q}$ can be fitted with a simple Arrhenius law.
}
\label{figure5}
\end{figure}

In the $\alpha$ regime,
we defined the relaxation time $\tau_{q}$
as the time it takes to the different correlators to decay from 1
to $1/e$. 
It is assumed that any characteristic times
belonging to the  $\alpha$ regime show asymptotically the same 
temperature dependence $\tau_{q} \sim \tau$.
The $\tau_{1}$ and $\tau_{2}$ relaxation times are defined
similarly for the orientational correlation functions $C_{l=1,2}$.
For a direct comparison, orientational
self intermediate scattering functions 
have been rescaled with the EISF as
$[F^{s}_{u}(t)-A_{0}(q)]/[1-A_{0}(q)]$
in order to decay with time from 1 to zero.
According to MCT, the $\alpha$ relaxation times
$\tau_{q}^{-1/\gamma}(T)$ should yield
straight lines intersecting the abscissa at $T = T_{c}$.
Fig.~\ref{figure5} shows that for all investigated
correlators this prediction holds well
over a relatively large temperature range.
Extrapolation of the temperature dependence
gives a critical temperature of $T_{c} \simeq 225 \pm 8 $ K
consistent with the $\beta$-regime analysis. About 40 K 
above $T_{c}$,
discrepancies with the MCT power law prediction 
are seen, as already shown in the fast regime,
and certainly due to hopping processes.
Furthermore, the time-temperature superposition principle
has been also checked and 
we found that all correlators fall onto a master curve
in the $\alpha$ time domain when
plotted as function of the rescaled time $t/\tau_{q}$ 
(not display in the present Letter).
However, at high temperatures, close to the dynamical crossover
previously observed at $T_{x}$ in~\cite{Affouard_epl},
a clear deviation of the $\alpha$ relaxation
from the MCT prediction is found. 
At $T > T_{x} \simeq 350 $ K, we have calculated that 
our data can be very well reproduced by a simple Arrhenius law.
The $T_{x}$ crossing temperature alters slightly for the different 
correlators. At present, we have no clear interpretation of this feature.
Nevertheless, as it has been nicely demonstrated by Sastry \emph{et al.}~\cite{Sastry,Sastry2}
for a binary Lennard-Jones MD simulation,
we obtained here for plastic crystal a strong indication of the existence 
of a 'landscape-influenced' regime occurring in the temperature range $[T_{c}-T_{x}]$. 
The dynamics should be 'landscape-dominated' only below $T_{c}$. 

In conclusion,
for the first time, this study
reveals that 
dynamics of one orientationally disorder crystal
can be relatively well described by 
the idealized version
of the MCT. It is noteworthy to remark that 
such an agreement was expected
for those systems where dynamics are almost completely
controlled by rotations, since 
some very recent extensions of MCT
to molecular systems~\cite{Chong,Fabbian,Winkler} 
have particularly shown that the asymptotic predictions of the idealized
version of MCT still hold.
For Cla plastic crystal, 
the same critical temperature $T_{c} \simeq 225 $ K
can be extracted in the $\alpha$ and $\beta$ regime
which strongly supports the MCT description.
Furthermore, this investigation
confirms the existence of a second remarkable dynamical crossover
at the temperature $T_{x} > T_{c}$ consistent with previous 
calorimetric~\cite{Oguni} and NMR and MD studies~\cite{Affouard_epl}.
This allows us to 
determine
precisely the temperature range $[T_{x} - T_{c}]$ of 
the influenced-landscape' regime in this rotator system
as defined recently by Sastry \emph{et al.}~\cite{Sastry,Sastry2}
for a model glass forming liquid.
Our results obtained for Cla call for new investigations:
(i) It will be of great 
interest to study the $q$ dependence of the different parameters 
$f^{c}_{q}$, $h^{(1)}_{q}$ or $\tau_{q}$ which has not been
performed in the present Letter for Cla due to 
the limited number of investigated wave vectors.
(ii) Experimental and numerical
extensions of the present study to other well documented 
orientationally disordered crystals
such as cyclooctanol or ethanol 
are needed to fully validate the MCT picture. 
(iii) 
Theoretical studies are highly desirable in order to clarify
the precise nature of the dynamical crossover occuring at $T_{x}$
which remains unclear at present.

\smallskip

The authors wish to acknowledge the use of the facilities of the IDRIS,
B\^atiment 506,
Facult\'e des Sciences d'Orsay F-91403 Orsay,
where some of the simulations were carried out.
This work was supported by the INTERREG II
program
(Nord Pas de Calais/Kent).



\begin{thebibliography}{10}
\bibitem{Affouard_cladm_model}
\bibinfo{journal}{Our system is made of molecules composed of two sites: one
  chlorine and one super atom that models the adamantane group
  $\mathrm{C_{10}H_{15}}$. Molecular dynamics calculations were performed on a
  system of $N = 256$ ($4 \times 4 \times 4$ fcc crystalline cells) molecules
  interacting through a short range site-site Lennard-Jones potential. The
  electrostatic interactions were handled by the Ewald method with two partial
  charges ($q= \pm 0.151$ e) localized on both sites which reproduce the
  experimental dipolar moment $|\vec{\mu}| \simeq 2.39 $ Debyes. A time step of
  $\Delta t = 5$ fs was used to integrate the Newton's equations of motion. We
  worked in the NPT statistical ensemble with periodic boundary conditions.} .
\end{thebibliography}
\end{document}